# Revisiting the Panko–Halverson Taxonomy of Spreadsheet Errors

Raymond R. Panko
University of Hawaii
Panko@hawaii.edu

## *1. Introduction*

The purpose of this paper is to revisit the Panko–Halverson [1996] taxonomy of spreadsheet errors and suggest revisions. There are several reasons for doing so.

- First, the taxonomy has been widely used. Therefore, it should have scrutiny.
- Second, the taxonomy has not widely available in its original form [Panko & Halverson, 1996]. Consequently, most users refer to secondary sources. Consequently, they often equate the taxonomy with the simplified extracts used in particular experiments or field studies.
- Third, perhaps as a consequence, most users use only a fraction of the taxonomy. In particular, they tend not to use the taxonomy's life-cycle dimension.
- Fourth, the taxonomy has been tested against spreadsheets in experiments and spreadsheets in operational use. It is time to review how it has fared in these tests.
- Fifth, the taxonomy was based on the types of spreadsheet errors that were known to the authors in the mid-1990s. Subsequent experience has shown that the taxonomy needs to be extended for situations beyond those original experiences.
- Sixth, the omission category in the taxonomy has proven to be too narrow.

Although this paper will focus on the Panko–Halverson taxonomy, this does not mean that that it is the only possible error taxonomy or even the best error taxonomy.

### *1.1 Taxonomies*

Taxonomies have long been used in science. Senders and Moray [1991], writing about human error, wrote that:

> … a taxonomy is a fundamental requirement for the foundation of empirical science. If we want a deep understanding of the nature, origins, and causes of human error, it is necessary to have an unambiguous classification scheme for describing the phenomenon we are studying. [Senders and Moray, 1991, p. 82.]

For our purposes, we will define a taxonomy as the division of a large number of entities into a number of related categories whose differences are useful for a particular purpose.

The first emphasis is the *ordering of many entities into categories*. Ideally, the categories will be comprehensive, encompassing all entities. In addition, the categories ideally should be mutually exclusive, without overlap. In mathematical terms, there should be a one-to-one correspondence between entities and categories.

The second emphasis is usefulness *for a particular purpose* [Senders and Moray, 1991]. There is no such thing as "best" error taxonomy for spreadsheets [Grossman and Özlük, 2003] or any other type of human cognitive activity. Researchers and professionals with different focuses may need different things from error taxonomies. For instance, designers need error taxonomies that distinguish between types of errors that need different amelioration strategies. The legal system, in contrast, needs distinctions that help assign responsibility for damages [Senders & Moray, 1991]. Researchers with different purposes need different things from taxonomies and so may need different taxonomies.

In addition, several taxonomies are needed because each taxonomy will illuminate some aspects of the phenomenon while blinding the researcher or practitioner to other aspects. This occurs because theories in general illuminate tend to some things while ignoring others. For example, Graham Allison and Philip Zelikow [1999] analyzed the Cuban missile crisis from the viewpoint of several different theories about decision making. They showed how each theory was shockingly oblivious to certain types of evidence.

### *1.2 Phenomenological versus Theory-Based Taxonomies*

Senders and Moray [1991] distinguished between different levels of taxonomy. The most superficial level consists of *phenomenological taxonomies* that are based on simple descriptions of error manifestations. For instance, typing errors at this level would be described by such things as key transpositions and other visible manifestation of errors. At the level of phenomenological errors, there is no explanation for why different errors occur, but taxonomies at this level may spur research into why specific types of errors occur.

Although one would prefer deeper taxonomies, phenomenological taxonomies can be very useful. Most obviously, they can focus subsequent research. In the human error field, if a particular certain type of error proves to be particularly frequent, then it may merit stronger attention. Conversely, if a type of error that was considered to be important actually turns out to be infrequent, then shifting resources to the study of other errors may be important.

Deeper taxonomies are informed by theory. This is especially valuable if theory predicts manifestations of results. In error research, for instance, theory may suggest different error occurrence rates for different types of errors, different detection rates, or different mechanisms for amelioration. Unfortunately, there is no complete theory for human error, so creating full deep taxonomies for spreadsheet errors is not possible.

Nearly all spreadsheet error research uses the post hoc analysis of spreadsheets that have already been developed. As a consequence, all of the error evidence is phenomenological. This would suggest that we should only be able to have phenomenological taxonomies. However, the human mind wants explanations. For better or worse, nearly all published taxonomies of error and spreadsheet error try to explain observed post hoc errors in terms of underlying theories, both formal and informal. While this may be fundamentally undesirable, it is also undesirable to use taxonomies that describe errors but give no clues as to why different types of errors occur or how they can be redressed.

### *1.3 Reliability in Classification*

Taxonomies, like any other research methodology can be judged on a number of methodological criteria. Every taxonomy should face the entire battery of tests required to assess its internal and external validity.

We will only mention one of these methodological issues, reliability. Reliability means that if different people use the taxonomy to classify the same events or items, they will classify individual items in the same way. A taxonomy that cannot be applied reliably by different people is a failed taxonomy.

The simplest way to test reliability is to conduct an inter-rater reliability study. In these studies, two (or preferably more) people conduct a classification, and their consistency is compared statistically. There are several statistical tests available for testing inter-rater reliability. In general, an inter-rater reliability of 90% or higher is the goal, although an inter-rater reliability of 70% may make a study publishable as an exploratory study.

In error analysis, using multiple raters has another benefit. It can allow us to estimate statistically how many errors remain undiscovered in the spreadsheet. However, this only works if a reliable error taxonomy is used because different types of errors have different detection rates. In general, ignoring error types will give an underestimate for the number of errors remaining, but even doing an estimate for number of error remaining based on all errors can be eye-opening for anyone who believes that they have found most or all errors in a spreadsheet.

## *2. General Human Error Taxonomies*

There have been many taxonomies of human errors. We cannot cover them all. However, we will cover a few that appear to be of particular importance for building a taxonomy of spreadsheet errors.

### *2.1 What is an Error?*

Senders and Moray [1991] defined an error as an action that is

> "not intended by the actor;
> not desired by a set of rules or an external observer; or
> that led the task or system outside its accepted limits"
> Senders and Moray (1991), p. 25.

Note that there needs to be an external standard for determining whether a result is an error. This can be general consensus that something is an error, or there can be a definitive test of whether a requirement has been satisfied or not. Not all errors have good external standards. This is particularly true for qualitative errors that violate good practice but do not, in the case of spreadsheets, generate an incorrect numerical answer.

Along with Senders and Moray [1991], we distinguish between errors and accidents. Accidents typically involve a series of errors and may even occur when no error has been made.

In programming, there is a distinction between faults and errors. A fault is a problem in the program. An error is a human action that leads to the fault. Most spreadsheet error taxonomies require the post-hoc analysis of spreadsheet models. For consistency, we could refer to problems found in post-hoc analysis as faults. However, outside of programming, this is not common terminology. We will use the term "error" as something that is incorrect in a spreadsheet model rather than as a human action that causes the problem.

### *2.2 Mistakes versus Slips and Lapses*

In his book, *Human Error*, Reason [1990] presented the taxonomy of human error types based on prior work by Reason and Mycielska [1982] and Norman [1981, 1984]. This taxonomy, which is shown in Figure 1, begins with a basic distinction between planning and implementation. If the plan is wrong, this is a mistake, regardless of how good the implementation is. However, if the plan is correct but the implementation is wrong, this is a slip or lapse.

*Figure 1: Mistakes versus Slips and Lapses*

The distinction between slips and lapses was proposed by Norman [1984]. A slip is an error during a sensory-motor action, such as typing the wrong number or pointing to the wrong cell. In contrast, a lapse occurs within the person's head. Typically, a lapse is a failure in memory, and this failure is often caused by overloading the limited human memory capacity.

This taxonomy has possible implications for automated spreadsheet analysis, which only works on final spreadsheet artifacts. It is likely that errors involving planning and storage that occur "off the spreadsheet" will leave few if any artifacts in the spreadsheet for automated analysis tools to find.

Even slips during execution may or may not leave artifacts for automated spreadsheet analysis programs to find.

For human error hunters, too, the three types of errors suggest that looking only at the spreadsheet is likely to miss many or even most errors. It is important to examine requirements, designs, and algorithms to understand if they have been executed properly in the spreadsheet.

### *2.3 Rasmussen and Jensen*

Rasmussen and Jenson [1974] observed highly experienced electricians doing troubleshooting. They used protocol analysis in order to understand why errors occurred. This provided data below the phenomenological errors.

They observed that the troubleshooters used different strategies to solving different types of problems. Figure 2 shows these three strategies.

*Figure 2: Rasmussen's Taxonomy of Cognition and Errors*

- Much of the time, the troubleshooters were simply applying sensory-motor skills, such as using a voltmeter.
- When this was not sufficient, they applied one of many rules they had learned over time. A typical rule might be, "First, check if the device is plugged in and turned on." While this rule is obvious, many rules are quite subtle and are developed only after years of practice.
- When no existing rules were applicable, the troubleshooters had to use their general knowledge about the specific devices being studied and about electronics in general.

This taxonomy is useful in expanding our understanding of the mental processes that come into play when people work and, therefore, how they make errors. Mistakes, in other words, can occur when doing rule-based thinking or knowledge-based thinking. Errors during rule-based and knowledge-based activities may be very different and may require different error reduction strategies.

Although the Rasmussen and Jensen typology is attractive, applying it tends to run into two significant problems. First, subjects must be experienced. Unless the subject is experienced, he or she is not likely to have developed many rules. In addition, if subjects are comparative novices, they may not yet have developed the understanding of the knowledge domain sufficient to allow them to do knowledge-based work. Due to these problems, any researcher who wishes to use the Rasmussen typology needs to use suitable subjects.

The second problem is that the Rasmussen and Jensen study observed work as it was being done. (It was a form of protocol analysis.) To use this taxonomy post hoc, based on observed errors in a completed spreadsheet, would require a great deal of justification, if such justification was possible at all. How is it possible, for example, to tell whether an activity was rule-based or knowledge-based when an error was concerned, simply on the basis of artifacts?

### *2.4 Allwood*

To build his taxonomy, Allwood [1984], did a protocol analysis with students solving mathematical problems. This is somewhat more specific than the other human error taxonomies we have seen because it only deals with mathematical errors. Of course, many spreadsheet errors are likely to be mathematical errors.

Figure 3 shows that Allwood's students made 327 errors as they worked. Six out of ten errors were execution errors, which involved something like doing a calculation incorrectly. The subjects spontaneously caught 83% of these errors as they worked. Consequently, execution errors accounted for only 29% of final errors.

*Figure 3: Allwood's Study of Mathematical Errors*

Errors that involved mathematical thinking, namely solution method errors and higher-level math errors, only accounted for a quarter of all errors made, but their relatively low error detection rates—48% and 25%, respectively, resulted in their accounting for 40% of all errors.

Skip errors involved subjects skipping a step in a solution process. These errors were comparatively rate, accounting for only 9% of all errors. However, none were detected spontaneously, so they resulted in 29% of all final errors.

Allwood [1984] showed that different types of errors in this taxonomy had radically different occurrence rates and detection rates. That certainly is attractive in error taxonomies. In addition, of course, spreadsheets generally perform mathematical computations. It may be possible to build on the Allwood framework to spreadsheet error taxonomies. Panko and Halverson [1996] did exactly that.

### *2.5 Flowers and Hayes*

Another limited but intriguing error taxonomy comes from Flower and Hayes [Flower and Hayes, 1980; Hayes and Flower, 1980], who studied the process of writing.

In another protocol analysis, Gould [1980] noted that writers spend 40% to 70% of their time thinking instead of actually writing. He found that they were reviewing what they had just written (similar to Allwood's [2004] standard check during in mathematics) and planning where they would go next. Gentner [1988] also noted that people spend time pausing when they are doing typing.

Flower and Hayes looked in depth at the non-writing time in the writing process. They found that their subjects had to work at several levels of abstraction simultaneously. They had to select specific words while generating sentences, and sentence production had to fit into the author's plan for the paragraph, for larger units of the document, and for the document as a whole. Planning had to be done at all levels of abstraction, and it had to be done simultaneously. Each level of abstraction created constraints that had to be obeyed when considering other levels.

Figure 4 shows that the Flower and Hayes taxonomy of concerns can be viewed as a context pyramid that is inverted, placing all of the weight of all context levels on the writing of a word. This can create enormous overload on the writer's memory and planning resources.

*Figure 4: The Flower and Hayes Context Pyramid*

In spreadsheet development, the same mental load is generated. Whenever a developer types a formula, he or she has to be cognizant of the algorithm for the formula, the algorithm for a larger section of the spreadsheet, and for the spreadsheet as a whole.

### *2.6 Jambon*

The final human error taxonomy we will mention was created by Jambon [1998]. In contrast to other taxonomies we have seen, the Jambon technology focuses on testing and remediation *after* development rather than during development. Jambon noted that testing and remediation is a fairly complex process involving two stages:

- Error diagnosis consists of error detection, followed by error explanation. At the end of this stage, the tester knows that a problem exists and what it is.
- Error recovery consists of the actions needed to fix an error. Jambon [1998] divided error recovery into planning an execution.

In addition to noting the complexity of testing and remediation, Jambon [1998] noted that there are two different approaches during error explanation. The first is forward error correction, which consists of doing things to get the correct results. The other consists of backward error detection, which consists of working from the error back to its cause.

### 2.7 Perspective on Human Error Taxonomies

The research that has been done to date on human error taxonomies suggests that human error is a complex process. The Norman–Reason taxonomy of mistakes, lapses, and slips appears to be very widely accepted and is backed by both theory and experimental data. However, each error taxonomy that we have seen (and many more) provides important insights into human error issues.

## 3. Spreadsheet Error Taxonomies

So far, we have been looking at general human error taxonomies. Building on this, we will now look specifically at spreadsheet error taxonomies.

### 3.1 Galletta

When Galletta et al. [1993] conducted an experiment in which MBA students and accountants working on their accreditation examined spreadsheets looking for errors, they divided errors into two types. First, there were *domain errors* that occurred when a formula required knowledge of accounting. Second, there were *device errors*, which occurred when the error involved using the computer and the spreadsheet program—typing errors and pointing errors.

### 3.2 Panko and Halverson

For their research on errors in spreadsheet development and inspection, Panko and Halverson [1996] created a taxonomy of spreadsheet research risks as a three dimensional cube. The three sides of this cube were research issue, life cycle stage, and methodology (experiment, survey, etc.) for addressing the research issues.

*Figure 5: Panko and Halverson Spreadsheet Risks Research Cube*

#### Research Issues

Research issues included structural concerns (poor structure), actual errors, user work practices, assumptions, and spreadsheet model characteristics (size, percentage of cells that are formulas or data, complexity of formulas, one-time versus many-time use, number of people who use the spreadsheet, purpose, and so forth), and control policies.

#### Measuring Error Rates

Under "actual errors," the taxonomy noted several ways to quantify errors and noted that each has advantages and disadvantages. The metrics listed were:

- Percentage of models containing errors
- Number of errors per model
- Distribution of errors by magnitude
- Cell error rate

*Figure 6: Panko and Halverson Metrics for Measuring Errors*

For error magnitude, Panko and Halverson noted that "Some errors are important, other unimportant. One measure is the size of the error as a percentage of the correct bottom-line value. Another is whether a decision would have been different had the error not been made. We suspect that quite a few errors are either too small to be important or still give answers that lead to the correct decisions." Many field audits have found that significant errors (such as errors that are material in financial statements or that can affect a decision) are very widespread but that "show stopper" errors only occur in about 5% of all spreadsheets.

In terms of the cell error rate, which is the percentage of cells that contain an error, Panko and Halverson were taking a cue from software development research, which has long measured the faults per thousand lines of noncomment source code (faults/KLOC). Within limits, the rate of faults/KLOC is reasonably the same across programs. This allows software developers to get a rough count of the number of errors they can expect to find when inspecting a module of code with known length.

In their taxonomy, Panko and Halverson did not discuss how to count errors. In the research in which this taxonomy was created, Panko and Halverson [1997, 2001] argued that errors should be counted in the cells in which they occurred. Even if this error is repeated in copied cells or makes dependent cells incorrect, it should only be counted as a single cell.

***Basic Error Types***

Figure 7 shows the Panko and Halverson [1996] taxonomy of error types. The taxonomy first divides errors into quantitative and quantitative errors. Their demarcation of the two types of errors was very simple. If something made a final value (bottom-line value) incorrect, then it was a quantitative error. If it did not, it was a qualitative error.

*Figure 7: Panko and Halverson Taxonomy of Error Types*

The most common qualitative error is putting a number into a formula instead of a cell reference. It does not cause errors then, but it makes errors more likely later, say when assumptions have to be changed in what-if analysis. In fact, Teo and Tan [1997] did find later that students who did jamming (hardcoding numbers in formulas) did make more errors in subsequent what-if analyses. Reason [1990] calls such errors latent errors because they cause no errors at the time they are made but increase the likelihood of an error occurring later. Pryor [2003] gives an excellent list of qualitative errors.

Following Allwood [1984] broadly, Panko and Halverson divided quantitative errors into three basic types: mechanical, logic, and omission errors. They have the following definitions for these error types:

- Mechanical errors are typing errors, pointing errors, and other simple slips. Mechanical errors can be frequent, but they have a high chance of being caught by the person making the error.
- Logic errors are incorrect formulas due to choosing the wrong Algorithm or creating the wrong formula to implement the algorithm.
- Omissions are things left out of the model that should be there. They often result from a misinterpretation of the situation. Human factors research has shown that omission errors are especially dangerous because they have low error rates.

Logic errors occurred frequently, and Panko and Halverson used two different ways to distinguish between them. First, Lorge and Solomon [1955] had talked about errors that are obvious when pointed out to the person who made the error. Lorge and Solomon called these Eureka errors. In their first study, Panko and Halverson observed students as they worked. They found that for certain errors, even if one of the team members warned about the error, he or she would often be ignored. The researchers called these Cassandra errors, after the Homeric character who was cursed to warn of disasters but never be believed. As Lorge and Solomon noted, groups are very good at reducing Eureka errors. Panko and Halverson showed that groups were very poor at reducing Cassandra errors.

Panko and Halverson also distinguished between pure logic errors and domain logic errors. Domain logic errors stemmed from the builder's misunderstanding of the knowledge domain for the spreadsheet, such as accounting. In contrast, pure logic errors resulted in the incorrect use of mathematics or logic in general. Panko created the Wall task, which was simple and free of domain knowledge requirements, in order to avoid the complications of requirements for domain knowledge. Panko and Sprague [1998] conducted an experiment using the Wall task. Subjects still made logic errors, indicating that errors in spreadsheets were not simply due to poor domain knowledge.

### *Using the Taxonomy*

Panko and Halverson's first study using the taxonomy was a development experiment in which subjects developed a spreadsheet working alone, in groups of two, or in groups of four [Panko and Halverson, 1997]. The authors conducted an inter-rater reliability test on the taxonomy's definition of quantitative errors (errors that changed a final value), the tripartite distinction between mechanical, logical, and omission errors, and the distinctions between Eureka and Cassandra errors. The subjects made the same 209 quantitative errors according to both researchers, for a 100% reliability rate. Within these quantitative errors, the researchers initially disagreed on the classification of a single error that occurred in three spreadsheets. This represented 99.6% reliability. The point of disagreement was a mistake made by three subjects who all added expenses to revenues to get income, instead of subtracting expenses. One researcher classified this as a logic error, the other as a mechanical error. After a discussion the authors agreed to call it a mechanical error. Teo and Tan [1997] reported no problems when they used the taxonomy in a duplicate of the Panko and Sprague [1998] Wall task.

Panko [1999] later conducted an inspection study, using a modification of the Galletta et al. [1997] inspection task. This time, Panko tested the distinction between omission errors and other types of errors (mechanical and logical). The data showed that omission errors were indeed detected much less frequently than other types of errors.

Hicks [1995] used the Panko [1999] inspection method and the Panko and Halverson [1996] error taxonomy to inspect a large capital budgeting spreadsheet about to be operational in a multi-billion dollar company. They used three inspectors working apart and then together to compare their results. (Unfortunately, they did not report the amount of time spent and the number of cells in the spreadsheet, contrary to the Panko and Halverson [1996, 1997] methodology. They reported that the taxonomy worked well for them, although they did not report inter-rater reliability.

### *Errors by Life Cycle*

The third dimension in the spreadsheet risks cubes was life cycle stage. Based on the prior spreadsheet literature, they divided the spreadsheet life cycle (not just the spreadsheet development life cycle) into 5 stages:

- Requirements and Design
- Cell Entry
- The Draft Stage (before testing)
- Debugging
- Operation

Panko and Halverson [1996] suggested that the error rate varies strongly across this life cycle. For early stages, they note that experiments indicate that when people enter formulas in cells, the error rate is about 10%, but that most of these errors are caught (Olson and Olson, 1990). Consequently, when people finish development (this is called the draft stage), the error rate is half of that or less. Third, when subjects inspect spreadsheets to look for errors, they find a majority of them, further reducing the error rate.

### *Problems with the Panko and Halverson Taxonomy*

Although the Panko and Halverson taxonomy has been fairly well validated by experiments, some limitations have become obvious over time.

- First, although the taxonomy has both a error type dimension and a spreadsheet life cycle perspective, this was not fleshed out until later. For instance, Panko and Halverson focused on development and inspection. They did not look at the types of errors that occur during initial analysis and requirements. More concretely, probably because they did not study ongoing use they were not aware until later of overwriting errors, in which a user overwrites a formula with a numerical value.

- Second, they focused on omission errors because these were the subject of earlier human error research. However, given the work of Flower and Hayes, omission errors are only one type of error that is likely to occur as people with limited memory resources must cope with simultaneous needs to enter a formula, keeping the full algorithm in mind, and keeping the flow of the entire spreadsheet in mind.
- Third, the taxonomy did not recognize the important distinction between sensorymotor slips and memory lapses. Particularly in the wall task [Panko and Sprague, 1998], dividing mechanical errors into slips and lapses would change the picture considerably.

### *3.3 Rajalingham*

Shortly after the Panko and Halverson [1996] taxonomy, Rajalingham, et al. [2000] created more complex taxonomy of spreadsheet errors. This taxonomy is shown in Figure 8.

*Figure 8: The Rajalingham, Chadwick, Knight, and Edwards 2000 Taxonomy*

This taxonomy also begins with the distinction between qualitative and quantitative errors. It then gets into the distinction between accidental errors and reasoning errors. This is similar to the Panko and Halverson [1996] mechanical versus logical distinction, but its terminology (accidental versus reasoning) may be better connotatively.

An important addition in this taxonomy is the distinction between developer and end-user errors. Panko and Halverson [1996] only focused on developer errors. They did not consider the types of errors that end users would make after development. Most obviously, they failed to consider data entry errors, which can be very important. These errors can include inputting incorrect data or even overwriting a number with a formula. Rajalingham, et al. [2000] also considers errors that users make in interpreting the results of spreadsheets. If a spreadsheet gives the correct result but this correct result is misinterpreted, say because of poor output labeling, this is just as bad as a development error.

Later, Rajalingham [2005] revisited the taxonomy. He actually came up with two follow-up taxonomies. Figure 9 shows his “bushy” taxonomy. He gave it this name because it often branches into three or more alternatives. Rajalingham argued that this approach made it difficult to decide where to place and error, and it also tended to require an error to be placed in two or more end nodes.

*Figure 9: Rajalingham's 2005 "Bushy" Taxonomy*

Rajalingham, in the same paper, also presented his “binary” taxonomy. He argued that having to make a binary choice at each step made classification easier and more predictable.

*Figure 10: Rajalingham's 2005 "Binary" Taxonomy*

### *3.4 Howe and Simkin*

In a code inspection experiment, Howe and Simkin [2006] created a new taxonomy of error. Figure 11 shows this taxonomy.

*Figure 11: Howe and Simkin Taxonomy*

Obviously, this taxonomy is very different from earlier taxonomy. Its clerical and non-material errors are such things as spelling errors in labels, incorrect dates, and so forth. Most previous studies ignored such errors.

There is an especially important addition in the rules violations category. These are basically parts of the model that violate requirements. Omission errors do this, but so do many other types of errors.

Giving evidence for the usefulness of the taxonomy, subjects had different detection rates for different types of errors. In another code inspection study, Bishop and McDaid [2007] used the Howe and Simkin [2006] taxonomy. They also found differences in error detection rates, and they found that experienced spreadsheet developers from industry had a higher detection rate than students for rules violations and formula errors. However, the Bishop and McDaid subjects had a far lower detection rate for clerical /nonmaterial errors, perhaps because they had not been explicitly instructed to look for them.

### *3.5 Powell, Lawson, and Baker*

For their series of projects involving the creation, testing, and use of a code inspection (auditing) methodology, Powell, Lawson, and Baker [2007] developed another taxonomy of errors. Figure 12 illustrates this taxonomy

*Figure 12: Powell, Lawson, and Baker Taxonomy*

Note that this taxonomy's use of omission errors is very different from the use of omission errors in Panko and Halverson [1996]. In the Panko and Halverson, something in the requirements is left out of the spreadsheet. This is not likely to be detectable by looking at the spreadsheet. In contrast, in the Powell, Lawson, and Baker [2007] taxonomy, it means pointing to a blank cell.

Hard coding is described as a qualitative error in the Panko and Halverson [1996] and the Rajalingham [2005] bushy taxonomy. In the Powell, Lawson, and Baker [2007] taxonomy, hard coding is usually not a quantitative error but sometimes is, "if it is sufficiently dangerous" (Page 60).

The spreadsheets studied to develop this taxonomy were operational spreadsheets in use for some time. However, there is no category for overwriting a formula with a constant. Nor is there any indication that this has happened.

### *3.6 Madahar, Cleary, and Ball*

In the EuSpRIG conference during which Powell, Lawson, and Baker [2007] presented their taxonomy, Madahar, Cleary, and Ball [2007] also presented their taxonomy. In contrast to other taxonomies, this was a taxonomy of spreadsheets rather than of error in the spreadsheet. Writers have long argued that different types of spreadsheets need greater or lesser degrees of control (e.g., Schultheis and Sumner, 1994).

Madahar, Cleary, and Ball [2007] considered three models for describing the different types of spreadsheets they found in the one department of a university. They described Model 3 as their best model. This model had three dimensions.

*Figure 13: Madahar, Cleary, and Ball Taxonomy of Spreadsheets*

- Dependency means how fundamentally the organization depends on the spreadsheet. Values can be operational, tactical, or strategic.
- Magnitude is the severity of consequences for potential errors.
- Time/Urgency refers to deadlines that have to be met using the spreadsheet.

## *4. Revising the Panko and Halverson Taxonomy*

Although the Panko and Halverson [1996] taxonomy has worked relatively well, it is more than a decade old and is showing its age. In particular, it has two obvious problems.

- First, it was developed for a specific purpose—to classify quantitative errors in spreadsheet development and inspection that reflects human differences in commission rate and detection rate. However, because it was limited in its approach, it did not consider errors that occur in other stages of the spreadsheet systems life cycle. In addition, because it was developed in an effort to prove that quantitative errors are in fact common and difficult to detect, it paid little attention to qualitative errors which are arguably more important.
- Second, even the taxonomy's view of quantitative errors was too limited. One specific problem is that its definition of mechanical errors included slips but not lapses. This is a serious problem. In addition, following Flower and Hayes [1980], the taxonomy was too focused at the cell level. However, errors can also occur if the developer loses focus on the broader flow of the spreadsheet.

### *4.1 Violations and Errors*

Figure 14 shows our revised taxonomy of violations and errors. It has many components

*Figure 14: Violations, Errors, and Context Levels*

#### *Violations versus Errors*

In software testing, Beizer's [1990] advice to hold developers blameless is widely followed. This reflects the realization that nobody is immune to error and that using testing to assign blame is counterproductive.

However, in the study of automobile accidents, researchers have long used a distinction between errors and violations. Violations are acts that break the law, such as speeding or driving while under the influence. While errors are inevitable and are not considered blameful, violations are considered blameful, even if they do not lead to accidents.

This distinction between errors and violations may be useful in spreadsheet development. In spreadsheet development, violations would normally consist of not complying with the organization's policies for spreadsheet development. Of course, this assumes that the organization is mature enough to have policies. In cases where the company is subject to external compliance regulations, then a violation would exist if a spreadsheet created a violation of the external compliance regulation.

#### *Qualitative versus Quantitative Errors*

This taxonomy continues to the distinction between qualitative and quantitative errors. Quantitative errors, quite simply, are incorrect formulas and data cells that cause subsequent dependent cells to have the wrong values. If an error does not cause a subsequent value to be wrong, then the error is not a quantitative error. It is a qualitative error. Nor is the issue seriousness of the error. Qualitative errors can be extremely serious.

#### *Mistakes, Slips, and Lapses*

Given the widespread use of the Reason and Norman distinction between mistakes, slips, and lapses, the taxonomy should be revised to reflect this set of categories.

For mistakes, it is important to realize that mistakes in formulas can come from many sources, including domain misunderstandings, logic failures, mathematical errors, and errors in using the software (usually by misusing built-in functions).

The impact of dividing the Panko and Halverson mechanical error category into slips and lapses is shown in Figure 15. In a corpus of spreadsheets described by Panko [2000], 82 subjects each developed a spreadsheet to provide a decision maker with a pro-forma income statement. In the corpus, 28% of the errors were logic errors, 21% were omission errors, and 41% were mechanical errors.

*Figure 15: Mechanical Errors, Slips, and Lapses*

The high percentage of mechanical errors was good news for error detection, because some mechanical errors such as pointing errors leave discoverable artifacts on the spreadsheet. However, the figure shows that when a classification based on slips and lapses is used, slips only account for 19% of the total errors, while 22% of the errors were lapses. Many of the lapses, by the way, occurred in reading requirements for unit costs for two years for labor and materials. This series of numbers seemed to create frequent overloads on the memory capacities of the developers.

### *4.2 Level of Analysis*

Flower and Hayes [1980] noted that developers constantly must take into account multiple levels of context. Figure 14 notes that the same is true in spreadsheet development. When someone works at the cell level, they also need to keep in mind what is happening at the algorithm level (most algorithms require a group of formula and data cells), at the module level, and at the level of the spreadsheet as a whole.

The developer also needs to take into account the entire business system in which the spreadsheet will be used. This includes management, organization, procedures, hardware, sources of data, and other matters.

### *4.3 Life Cycle Stages and Roles*

Spreadsheets generally go through a system life cycle that begins with the analysis of the current situation and needs and ends when the spreadsheet is terminated or replaced. Figure 16 shows the main stages in the system life cycle.

*Figure 16: The Spreadsheet Life Cycle and Types of Errors*

The first part of this life cycle is the system development life cycle, which includes initial analysis, the specification of requirements, the development of modules, the development of the full spreadsheet (by combining modules), and implementation. However, most of a spreadsheet's life is spent in operational use, and maintenance also has to be done occasionally. Finally, the spreadsheet is replaced or simply terminated.

The original Panko and Halverson [1996] taxonomy noted that the number of errors typically varies over a spreadsheet's life cycle. During development, many errors will exist. However, testing, inspection and use experience tend to reduce the number of errors during development. By the time a spreadsheet is released for use, good practice should substantially decrease the number of residual errors. During operational use, however, errors may increase if people if people input incorrect data or overwrite formulas with numbers.

More importantly, as Figure 16 attempt to illustrate, different types of errors will occur at different stages of the systems life cycle. Panko and Halverson [1996] only focused on development and testing. Consequently, they only considered the errors that occur during that stage of the spreadsheet life cycle.

Although the main violation and error categories are likely to occur over the entire life cycle, their specific manifestations will be very different at each stage. One of the main jobs of spreadsheet researchers must be to enumerate the kinds of errors that can and do exist at each stage.

Arguably the most important stage to understand is operational use. Many specific errors, such as entering the wrong number for a variable or incorrectly importing data, occur primarily during operational use. Violations also must be anticipated, such as violations of privacy or the use of spreadsheets to commit fraud.

Figure 17 shows another aspect of life cycle thinking. This is the fact that there are several possible organization roles involved. We need to think about violations and errors for each of these roles during each stage of the life cycle. Although these roles may be combined in many cases, it may still make sense to think in terms of logical roles to envision errors.

*Figure 17: Life Cycle Stages and Roles*

### *Perspective*

This paper has considerably revised and expanded the Panko and Halverson [1996] taxonomy of spreadsheet errors. The purpose of that taxonomy was to support quantitative research studies to demonstrate that quantitative spreadsheet errors are frequent, that quantitative spreadsheet errors are difficult to detect, and that many spreadsheet errors are significant.

To some extent, these ideas have been broadly accepted. In any case, people who still reject that experimental evidence regarding them are not likely to have their opinions changed by further quantitative research.

It is now time to shift our focus toward identifying the large number of different types of errors that are possible in different life cycle stages and by people with different roles to play. For this, we do not need tight taxonomies as much as we did previously. Although experiments and other quantitative research must have well-formed and tight logical taxonomies, as we move from proof or danger to providing guidance to corporations, we will need more expansive taxonomies that suggest issues rather than tight taxonomies to confirm issues.

## *References*

Allison, Graham T. and Zelikow, Phillip. (1999). *Essence of Decision: Explaining the Cuban Missile Crisis* (2nd Edition) (Paperback), Longman Publishers: Englewood Cliffs, NJ.

Allwood, C. M. (1984). "Error Detection Processes in Statistical Problem Solving." *Cognitive Science,* 8(4), 413-437.

Ayalew, Yirsaw; Clermont, Markus; & Mittermeir, Roland T. (2000, July 17-18). "Detecting Errors in Spreadsheets," *Symposium Proceedings EuSpRIG 2000*, University of Greenwich, London, UK, European Spreadsheet Risks Interest Group, 51-62.

Beizer, B. (1990). *Software Testing Techniques*. 2nd ed. New York: Van Nostrand.

Bishop, Brian and McDaid, Kevin, "An Empirical Study of End-User Behaviour in Spreadsheet Error Detection and Correction," *Proceedings of the European Spreadsheet Risks Interest Group, EuSpRIG 2007 Conference*, University of Greenwich, London, July 2007, pp. 165-176.

Croll, G. The importance and criticality of spreadsheets in the City of London. *Proceedings of the EuSpRIG Conference*, London (2005).

Flower, L. A., & Hayes, J. R. (1980). "The Dynamics of Composing: Making Plans and Juggling Constraints," *Cognitive Processes in Writing*. Eds. L. W. Gregg & E. R. Steinberg. Hillsdale, NJ: Lawrence Erlbaum Associates. 31-50.

Galletta, D. F.; Abraham, D.; El Louadi, M.; Lekse, W.; Pollailis, Y.A.; & Sampler, J.L. (1993, April-June). "An Empirical Study of Spreadsheet Error-Finding Performance." *Journal of Accounting, Management, and Information Technology, 3(2),* 79-95.

Gentner, D. R. (1988) "Expertise in Typewriting" in Chi, M. T. H., R. Glaser, and M. J. Faar (eds.) (1988), *The Nature of Expertise,* Hillsdale, NJ: Lawrence Erlbaum Associates, pp. 1-22.

Gould, John D. "Experiments on Composing Letters: Some Facts, Some Myths, and Some Observations, Chapter 5 in Lee W. Gregg and Erwin Steinberg (eds.) *Cognitive Processes in Writing*, Lawrence Erlbaum: Hillsdale, NJ, 1980, pp. 97-127.

Grossman, Thomas A. and Özlük, Özgür, "Research Strategy and Scoping Survey on Research Practices," *Proceedings of EuSpRIG 2003*, European Spreadsheet Risks Interest Group, July 24-25, Trinity College, Dublin, Ireland, pp. 23-32.

Hayes, J. R. & Flower, L. (1980). "Identifying the Organization of Writing Processes," *Cognitive Processes in Writing*. Eds. L. W. Gregg & E. R. Steinberg. Hillsdale NJ: Erlbaum. 31-50.

Howe, Harry & Simkin, Mark F. (2006, January). "Factors Affecting the Ability to Detect Spreadsheet Errors," *Decision Sciences Journal of Innovative Education*, *4(1),* 101-122.

Jambon, Francis. "Taxonomy for Human Error and System Fault Recovery from the Engineering Perspective," Proceedings of the *International Conference on Human–Computer Interaction in Aeronomics*, Montreal, Canada, May 1998, 55-60.

Madahar, Mukul; Cleary, Pat; and Ball, David. (2007). "Categorisation of Spreadsheet Use within Organisations, Incorporating Risk: A Progress Report," Proceedings of the European Spreadsheet Risks Interest Group, EuSpRIG 2007 Conference, University of Greenwich, London, July 2007, pp. 37-45.

Norman, Donald A., "Categorization of Action Slips," *Psychological Review*, 88, 1981, 1-15.

Panko, R. R. (2008a). Human Error Website. (http://panko.shidler.hawaii.edu/humanerr.htm). Honolulu, HI: University of Hawai`i.

Panko, R. R. (2008b). Spreadsheet Research (SSR) Website. (http://panko.shilder.hawaii.edu/panko/ssr/). Honolulu, HI: University of Hawai`i.

Panko, Raymond R., "Applying Code Inspection to Spreadsheet Testing," *Journal of Management Information Systems*, *16(2),* Fall 1999, 159-176.

Panko, Raymond R. & Sprague, Ralph H., Jr. (1998, April) "Hitting the Wall: Errors in Developing and Code Inspecting a 'Simple' Spreadsheet Model," *Decision Support Systems*, *22(4),* 337-353.

Panko, R. R. (1988). *End User Computing: Management, Applications, and Technology*. New York: Wiley.

Panko, Raymond R. and Halverson, Richard P., Jr., "An Experiment in Collaborative Spreadsheet Development," 2(4) *Journal of the Association for Information Systems*, July 2001.Panko, Raymond R. & Halverson, Richard Jr., "An Experiment in Team Development to Reduce Spreadsheet Errors" *Journal of Management Information Systems 15*(1), Spring 1997, 21-32

Panko, Raymond R. & Halverson, Richard Jr., "Are Two Heads Better than One? (At Reducing Errors in Spreadsheet Modeling)," *Office Systems Research Journal 15*(1), Spring 1997, 21-32.

Panko, Raymond R. and Halverson, R. H., Jr. "Spreadsheets on Trial: A Framework for Research on Spreadsheet Risks," *Proceedings of the Twenty-Ninth Hawaii International Conference on System Sciences*, *Volume II*, Kihei, Maui, January, 1996, pp. 326-335.

Powell, Stephen G.; Lawson, Barry; and Baker, Kenneth R. (2007). "Impact of Errors on Operational Spreadsheets," *Proceedings of the European Spreadsheet Risks Interest Group, EuSpRIG 2007 Conference*, University of Greenwich, London, July 2007, pp. 57-68.

Pryor, Louise, "Correctness is not Enough," *Proceedings of EuSpRIG 2003*, European Spreadsheet Risks Interest Group, July 24-25, Trinity College, Dublin, Ireland, pp. 117-122.

Rajalingham, Kamalasen; Chadwick, David; Knight, Brian; and Edwards, Dilwyn, (2000a, January)"Quality Control in Spreadsheets: A Software Engineering-Based Approach to Spreadsheet Development," Proceedings of the Thirty-Third Hawaii International Conference on System Sciences, Maui, Hawaii. (.pdf format)

Rajalingham, Kamalasen; Chadwick, David R.; & Knight, Brian. (2000b, July 17-18). "Classification of Spreadsheet Errors, *Symposium Proceedings EuSpRIG 2000*, University of Greenwich, London, UK, *European Spreadsheet Risks Interest Group*, pp. 23-34.

Rajalingham, Kamalasen. (2005, July). "A Revised Classification of Spreadsheet Errors," *Proceedings of the 2005 European Spreadsheet Risks Interest Group, EuSpRIG 2005*, Greenwich, London, 185-199.

Rasmussen, J. & Jensen, A. "Mental Procedures in Real-Life Tasks: A Case Study of Electronic Troubleshooting," *Ergonomics*, 1974, 17, 293-307.

Reason, James P. (1990). *Human Error*, Cambridge University Press: Cambridge, England, 1990.

Reason, James T. & Mycielska, K., *Absent-Minded? The Psychology of Mental Lapses and Everyday Errors*, Prentice Hall, Englewood Cliffs, N.J., 1982.

Senders, John W. and Moray, Neville P. (1991). *Human Error: Cause, Prediction, and Reduction*, LawrenceErlbaum: Hillsdale, NH.

Teo, T.S.H. & Tan, M., "Quantitative and Qualitative Errors in Spreadsheet Development," *Proceedings of the Thirtieth Hawaii International Conference on System Sciences, Kihei, Hawaii*, January 1997.

## *Figures*

*Figure 1: Mistakes versus Slips and Lapses*

| Stage of Error | Type of Error | |
|---|---|---|
| Error in Planning | Mistake | Logic or mathematical error, etc. |
| Error in Execution | Slip | Sensory-motor error |
| | Lapse | Error cause by memory overload |

*Figure 2: Rasmussen's Taxonomy of Cognition and Errors*

| Type of Work | Characteristics |
|---|---|
| Knowledge-Based | Applies knowledge of the system when no rules exist. |
| Rule-Based | Applies rules. Example: Things work better when they are plugged in and turned on. |
| Skill-based | Applies well-learned sensory-motor skills. Example: Measuring a voltage on a volt meter. |

*Figure 3: Allwood's Study of Mathematical Errors*

| | Execution | Solution Method | Higher level math | Skip | Other | Total |
|---|---|---|---|---|---|---|
| Total Errors | 202 | 67 | 16 | 29 | 13 | 327 |
| % of Errors Made | 62% | 20% | 5% | 9% | 4% | 100% |
| Found | 168 | 32 | 4 | 0 | 6 | 210 |
| % Found | 83% | 48% | 25% | 0% | 46% | 64% |
| Not Found | 34 | 35 | 12 | 29 | 7 | 117 |
| % of Final Errors | 29% | 30% | 10% | 25% | 6% | 100% |

*Figure 4: The Flower and Hayes Context Pyramid*

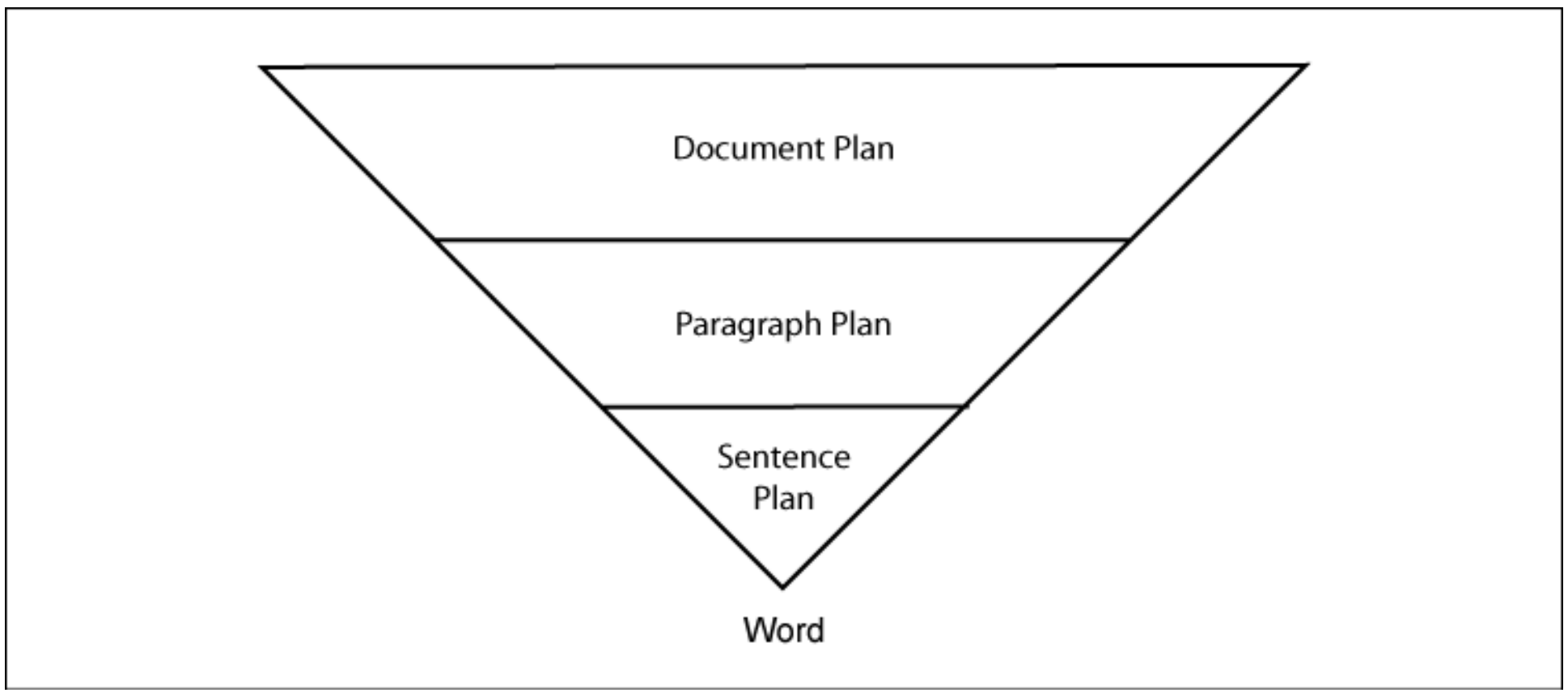

*Figure 5: Panko and Halverson Spreadsheet Risks Research Cube*

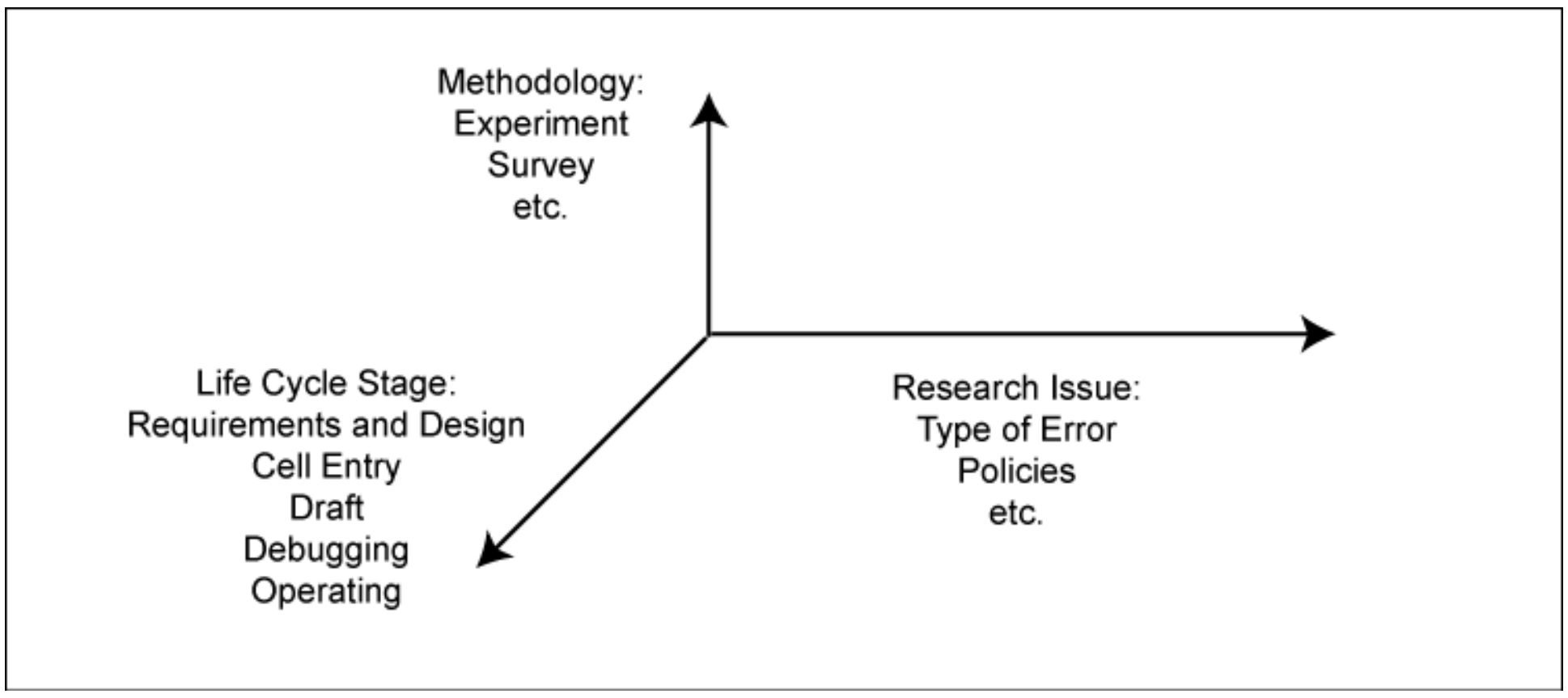

*Figure 6: Panko and Halverson Metrics for Measuring Errors*

Percentage of models containing errors
Number of errors per model
Distribution of errors by magnitude
Cell error rate

Note: Errors are recorded in the cell in which they originally occur. Consequent inaccuracies in copied cells or descendent cells due to this error are not counted as errors.

*Figure 7: Panko and Halverson Taxonomy of Error Types*

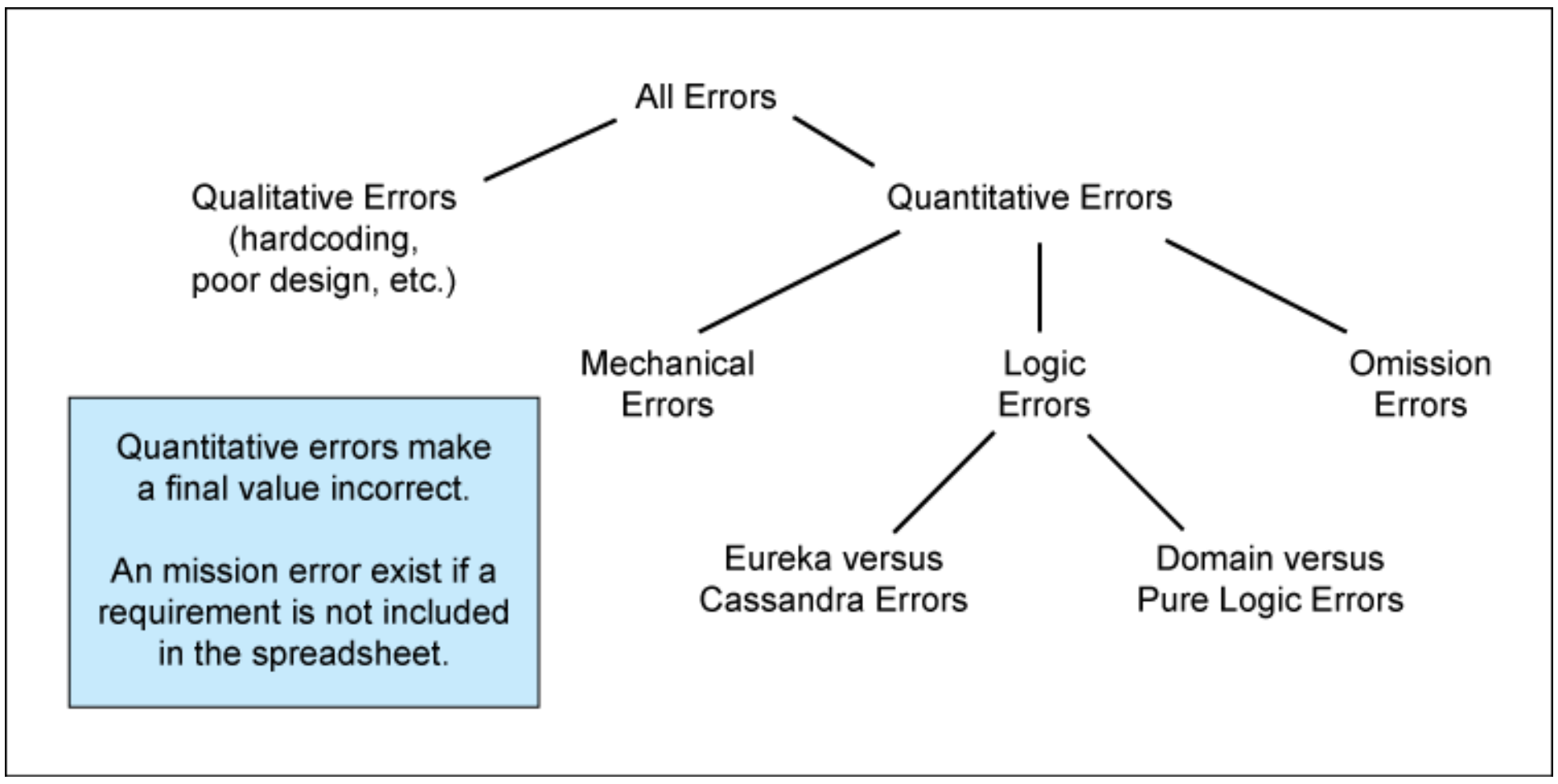

*Figure 8: The Rajalingham, Chadwick, Knight, and Edwards 2000 Taxonomy*

- **System-Generated**
- **User-Generated**
  - **Quantitative**
    - Accidental
      - Developer
        - Omission
        - Alteration (makes an incorrect change)
        - Duplication
      - End-User
        - Data inputter (input)
          - Omission
          - Alteration
          - Duplication
        - Interpreter (output)
          - Omission
          - Alteration
          - Duplication
    - Reasoning
      - Domain Knowledge
        - Real-world knowledge
        - Mathematical representation
      - Implementation
        - Syntax
        - Logic
  - **Qualitative**
    - Semantic
      - Structural
      - Temporal (based on information that has not been updated)
    - Maintainability

*Figure 9: Rajalingham's 2005 "Bushy" Taxonomy*

**Software Errors**

**User Errors**

- Qualitative Errors
  - Formatting errors
  - Update errors
  - Hard-coding errors
  - Semantic errors
- Quantitative Errors
  - Mechanical Errors
    - Overwriting Errors
      - Unreferenced data
      - Referenced data
    - Data Input Errors
      - Unreferenced data
      - Referenced data
  - Logic Errors
    - Errors in enabling skills
    - Errors in planning skills
  - Omission Errors

*Figure 10: Rajalingham's 2005 "Binary" Taxonomy*

**Quantitative**

- Accidental
  - Structural
    - Insertion
    - Update
      - Modification
      - Deletion
  - Data input
    - Insertion
    - Update
      - Deletion
      - Modification
- Reasoning
  - Domain knowledge
    - Real-world knowledge
    - Mathematical representation
  - Implementation
    - Logic
    - Syntax

**Qualitative**

- Temporal
- Structural
  - Visible
  - Hidden

*Figure 11: Howe and Simkin Taxonomy*

| Type of Error | Seeded Errors | Percentage Found | Description |
|---|---|---|---|
| Data Entry Errors | 5 | 72% | Out of range values, negative values, one value entered as a label |
| Clerical and Non-Material Errors | 10 | 66% | Spelling errors, incorrect dates, etc. |
| Rules Violations | 3 | 60% | Cell entries which violate a stated company policy for an ineligible employee |
| Formula Errors | 25 | 54% | Inaccurate range references, embedded constants, illogical formulas |
| Total Errors | 43 | 67% | |

*Figure 12: Powell, Lawson, and Baker Taxonomy*

| Error Type | Description |
|---|---|
| Logic | Formula is used incorrectly, leading to an incorrect result. |
| Reference | A formula contains one or more incorrect references to other cells. |
| Hard-Coding | One or more numbers appear in formulas, and the practice is sufficiently dangerous. |
| Copy/Paste | A formula is wrong do to an incorrect cut and paste. |
| Data Input | An incorrect data input is used. |
| Omission | A formula is wrong because one of its input cells is blank. |

*Figure 13: Madahar, Cleary, and Ball Taxonomy of Spreadsheets*

| Dimension | Description |
|---|---|
| Dependency | How fundamentally the organization depends on the spreadsheet. Values can be operational, tactical, or strategic |
| Magnitude | The severity of consequences for potential errors |
| Time/Urgency | Deadlines that have to be met using the spreadsheet |

*Figure 14: Violations, Errors, and Context Levels*

| | Violations | Errors | | | | | | | |
|---|---|---|---|---|---|---|---|---|---|
| | | Qualitative | Quantitative | | | | | | |
| | | | Mistakes | | | | Slips/Lapses | | |
| Level | | | Domain | Logic | Math | Software | Slips | Lapses | |
| Business System | | | | | | | | | |
| Spreadsheet | | | | | | | | | |
| Module | | | | | | | | | |
| Algorithm | | | | | | | | | |
| Cell | | | | | | | | | |

*Figure 15: Mechanical Errors, Slips, and Lapses*

| | 1996 Taxonomy | Modification | |
|---|---|---|---|
| Type of Error | Mechanical | Slip | Lapse |
| Pointing errors | 8 | 8 | 0 |
| Year 1 and Year 2 sales salaries translated into two salespeople instead of two years | 1 | | 1 |
| Owner salary = 60,000 instead of 80,000 | 1 | | 1 |
| Typing incorrect value for unit materials and labor cost (usually due to a transposition) | 12 | | 12 |
| Units sold value for Year 2 used in Year 1 | 1 | | 1 |
| Units sold value 32,000 instead of 3,200* | 1 | 1 | |
| Sign incorrect | 2 | 2 | |
| Parenthesis error | 1 | 1 | |
| Rent = 3,600 instead of 36,000* | 1 | 1 | |
| Total Mechanical/Slip/Lapse Errors | 28 | 13 | 15 |
| Percentage of errors | 41% | 19% | 22% |

Note: * means could be categorized either as a slip or as a lapse.

*Figure 16: The Spreadsheet Life Cycle and Types of Errors*

| | Violations | Qualitative Errors | Mistakes | Slips and Lapses |
|---|---|---|---|---|
| Analysis | | | | |
| Requirements Development | | | | |
| Module Development | | | | |
| Spreadsheet Development | | | | |
| Implementation | | | | |
| Operation | | | | |
| Maintenance | | | | |
| Termination/Replacement | | | | |

*Figure 17: Life Cycle Stages and Roles*

| | Development | | | Operators | | |
|---|---|---|---|---|---|---|
| | Manager | Developer | Tester | Owner | Customer | Operator |
| Analysis | | | | | | |
| Requirements Development | | | | | | |
| Module Developments | | | | | | |
| Spreadsheet Development | | | | | | |
| Implementation | | | | | | |
| Operation | | | | | | |
| Maintenance | | | | | | |
| Termination/Replacement | | | | | | |